\pdfoutput=1
\documentclass[modern]{aastex61}

\received{\today}
\revised{\today}
\accepted{\today}

\submitjournal{ApJ}

\newcommand{\beq}{\begin{equation}}
\newcommand{\eeq}{\end{equation}}

\newcommand{\kms}{\mbox{ km s$^{-1}$}~}

\newcommand{\Mo}{\mbox{M$_{\odot}$}}
\newcommand{\Moy}{\mbox{M$_{\odot}$ yr$^{-1}$}~}

\newcommand{\Ro}{\mbox{R$_{\odot}$}}

\shorttitle{Common Envelope Shaping of Planetary Nebulae. II.}

\shortauthors{Garc\'{\i}a-Segura et al.}

\begin{document}

\title{Common Envelope Shaping of Planetary Nebulae. II. Magnetic Solutions and  Self-Collimated Outflows}

\correspondingauthor{Guillermo Garc\'{\i}a-Segura}
\email{ggs@astrosen.unam.mx}

\author{Guillermo Garc\'{\i}a-Segura}
\affiliation{Instituto de Astronom\'{\i}a, Universidad Nacional Aut\'onoma
de M\'exico, Km. 107 Carr. Tijuana-Ensenada, 22860, Ensenada, B. C., Mexico}

\author{Ronald E. Taam}
\affiliation{Center for Interdisciplinary Exploration and Research in Astrophysics (CIERA), Department of Physics and Astronomy, 
Northwestern University, 2145 Sheridan Road, Evanston, IL 60208, USA}

\author{Paul M. Ricker}
\affiliation{Department of Astronomy, University of Illinois, 1002 W. Green St., Urbana, IL 61801, USA}

\begin{abstract}

Magnetic fields of order $10^1-10^2$ gauss that are present in the envelopes of red giant stars are ejected in common 
envelope scenarios. These fields could be responsible for the launching of magnetically driven winds in proto-planetary nebulae. Using 2D simulations of magnetized winds interacting with an envelope drawn from a 3D simulation of the common envelope phase, we study the confinement, heating, and magnetic field development of post-common envelope winds. We find that the ejected
magnetic field can be enhanced via compression by factors up to $\sim 10^4$ in circumbinary disks during the self-regulated phases. We find values for the kinetic energy of the order of $10^{46}$ erg that explain the large values inferred in proto-planetary nebula outflows. We show that the interaction of the formed circumbinary disk with a spherical, stellar wind produces a ``tapered'' flow that is
almost indistinguishable from an imposed tapered flow. This increases the uncertainty of the origin of proto-planetary nebula winds, which could be either stellar, circumstellar (stellar accretion disk), circumbinary (circumbinary  accretion disk), or a combination of all three.

Within this framework, a scenario for self-collimation of weakly magnetized winds is discussed,
which can explain the two objects where the collimation process is observationally resolved, 
HD 101584 and Hen 3-1475.  An explanation for the equatorial, molecular hydrogen emission in CRL 2688 is also presented.

\end{abstract}

\keywords{ Stars: Evolution ---Stars: Rotation ---Stars: RG, AGB and 
post-AGB ---Stars: binaries: general  ---ISM: planetary nebulae: general---ISM: individual
(HD 101584, Hen3-1475, CRL 2688, MyCn 18)}

\section{Introduction} 
\label{sec:intro}

In our previous article (Garc\'{\i}a-Segura et al. 2018; paper I) we explicitly showed that during a common envelope (CE) 
event, an unbound part of the stellar envelope of the giant star is launched and ejected towards the circumstellar medium. 
The bound part of it falls back, creating an accretion disk, if a merger is produced, or else a circumbinary accretion disk, if 
the stellar companion survives, producing a short-period binary system (Ricker \& Taam 2012; Ivanova et al. 2013). The 
disks in both cases could, in principle, produce magnetically driven winds and jets (Soker \& Livio 1994; Romanova \& Owocki 2015).
Moreover, parts of a circumbinary accretion disk will fall into both stars, probably forming two smaller accretion disks. 
Taking into account that the remnant of the primary star is hot and able to form a line-driven wind, what we have to study in 
this scenario is a whole ``wind soup,'' which is the sum of all winds. This total outflowing  wind has been already assumed and modeled 
as a ``tapered'' flow  by Lee \& Sahai (2003) and more recently by Balick et al. (2019). One of the main topics of this article 
is to show that this total wind behaves, indeed, as a tapered flow due to the presence of a circumbinary disk, which acts as an obstacle.  A tapered flow is a
flow that has a density and velocity distribution that declines with the polar angle. However, it could also be described as a jet in a general context. Thus, a tapered flow is just a specific class of jet.

The origin of the magnetic field necessary for the magnetic shaping of planetary nebulae (PNe) (Chevalier \& Luo  1994; R\'o\.zyczka 
\& Franco 1996; Garc\'{\i}a-Segura 1997; Garc\'{\i}a-Segura et al. 1999; Garc\'{\i}a-Segura et al. 2005; Garc\'{\i}a-Segura et al. 2014) 
has been a major issue in the field of planetary nebulae. The discussion has been centered on the question of whether the stellar surface magnetic 
fields are sufficiently strong to become dynamically important, and how long this field can be maintained (Soker 2006). However,
this issue is now moot 
in the context of CE events, since the magnetic field that is important and relevant is the one in the inner 
envelope and not the one at the surface, as we show in this article. The field in the interiors of evolved stars is produced mainly at the 
core-envelope boundary, due to a large shear (Spruit 2002; Garc\'{\i}a-Segura et al. 2014). Thus, when the envelope is suddenly ejected in 
the spiral-in phase, the frozen-in magnetic field is carried out as well, with part of this field falling back and collapsing onto the circumbinary 
disk. The question now is if this field is dynamically important to produce magnetically driven winds. In this article we show that the 
field  is enhanced by a factor of $\sim 25$ due to compression when the
bound part of the envelope collapses to form the circumbinary disk, 
although it could be further amplified if a dynamo operates on the disk or during the CE event (Ohlmann et al. 2016). 

This article also investigates the self-magnetic collimation of the winds, and we are able to reproduce in general terms two observed objects, 
HD 101584 (Olofsson et al. 2019) observed with the Atacama Large Millimeter Array (ALMA)  and 
Hen 3-1475 (Fang et al. 2018) observed with the 
Hubble Space Telescope (HST), in which the process of self-collimation can be directly observed. 

The article is a continuation of paper I and is structured as follows: the numerical scheme and physical approximations, as well as the inputs for our calculations, are described in \S~2.  The results of the numerical simulations are presented in \S~3.   Finally, we discuss the numerical results in \S~4 and provide the main conclusions in the last section.

\section{Physical assumptions and numerical methods}

\subsection{Numerical simulations}

The simulations have been performed using the magneto-hydrodynamic code
ZEUS-3D (version 3.4), developed by M. L. Norman and the Laboratory for
Computational Astrophysics (Stone \& Norman 1992; Clarke 1996),
as in paper I, though the simulations there were purely hydrodynamical.

The computational grids consists of $200 \times 200$ (Models A and B) and $400 \times 400$ (Model C)
equidistant zones in spherical coordinates $r$ and $\theta$ respectively, with an angular extent of $90^{\circ}$, and an
initial radial extent of $1.9 \times 10^{13}$ cm (= 1.266 AU). The radial resolution is 
$9.5 \times 10^{10} $ cm (Models A and B) and $4.75 \times 10^{10}$ cm (Model C). 
Models A and B have fixed grids during the evolution with outflow outer boundary conditions. Model C
uses the self-expanding grid technique as in paper I, with inflow outer boundary 
conditions, which include a correction for the wind-compressed zone according to Bjorkman \& Casinnelli (1993)  
and Garc\'{\i}a-Segura et al. (1999) (see paper I).

\subsection{The initial conditions}

We use the same input conditions as in paper I, namely the numerical data describing the outcome of the CE computations 
presented in Ricker \& Taam (2012) using the FLASH code. The input data are originally three-dimensional in Cartesian coordinates, 
but our computations presented in this paper are performed in two spatial dimensions ($r,\theta$). The 3D densities and temperatures  
are averaged over $\phi$, and the velocity components ($v_r, v_{\theta}, v_{\phi}$) are averaged over $\phi$ using density weighting. 
2D spherical coordinates are very convenient since they allow the use of homologous, expanding grids in the radial direction.
Figure 1 illustrates the averaged density distribution on the left at 56.7 days of the CE calculation, which shows the ejected envelope 
and represents the initial condition for this study. 

The CE calculation in Ricker \& Taam (2012) does not include magnetic fields, but we do include them here in 
an approximation where they are not dynamically important. In order to assign magnetic field values to the stellar ejecta, 
we have first studied the magnetic fields generated in the stellar interior of two cases using the Binary Evolution Code (Petrovic et al. 2005; Yoon et al. 2006). The first one is Model r5 of Garc\'{\i}a-Segura et al.(2016). This is a model with an initial mass of
2.5 \Mo \,  and initial rotation of 250 \kms on the zero-age main sequence (ZAMS), and it is studied when the asymptotic giant branch 
(AGB) star has a radius of 415 \Ro \, at the beginning of the Roche lobe overflow (RLOF) phase. The second one is a new model for 1.1 \Mo \, and rotational velocity of  50 \kms on the ZAMS, and it is studied when the red giant (RG) star has a radius of 30 \Ro\, (similar to the case studied in Ricker \& Taam 2012). 
These two models predict that the toroidal magnetic field (the dominant field component) in the stellar interiors is of the order of
$10^4$ gauss  at the inner envelope and of the order of $10^1-10^2$ gauss near the surface. With that in mind, we have assumed that
the ejected envelope has a frozen-in toroidal field equal to $0.05 \times B_{\rm equi}$, which is a field that is not dynamically 
important and is far from the equipartition value $B_{\rm equi}$. To be conservative and speed up the calculation, we impose a 
maximum value of $500$ gauss on the grid
for the initial conditions. 
Otherwise, the field would reach $16,200$ gauss at the center. The resulting magnetic
field is shown on the right of Figure 1.

\subsection{The magnetized outflow}

As mentioned in the introduction, the post-CE outflow  should be a sum of several winds (a stellar line-driven wind plus 
one or several disk winds), and is treated here as a source of mass and momentum at the inner, radial boundary condition. We do 
not know the exact amount of the sum of all winds, but at least for now, we have an estimate for the line-driven wind of the 
hot stellar component of 0.36 \Mo  (core of RG phase),  descended from an initial mass of 1.05 \Mo. This core is a factor 
$\sim  0.03$ lower in luminosity than the core mass of 0.569 \Mo\, at the post-AGB phase computed by 
Vassiliadis \& Wood (1994), for an initial 1 \Mo \, stellar model. 
Using the analytical expressions in Villaver et al.(2002), 
we obtain a mass-loss rate of $4.19 \times 10^{-10} $ \Moy for the hot core, which is a truncated post-RG star, with a wind velocity of
630 \kms, $T_{\rm eff} = 29,000 $ K , and $5.1 \times 10^{45}$  ionizing photons per second. 
In this paper we focus on the post-RG case and reserve the post-AGB cases of paper I (Models A1 and A2) for a future study. 
The reason is that HD 101584 is probably a post-RG nebula (Olofsson et al. 2019), which probably matches with the case computed by 
Ricker \& Taam (2012), at least in order of magnitude. 
According to Reimers (1975), the RG star has a mass-loss rate of $\sim 10^{-9}$ \Moy previous to the CE phase.

We assume that the magnetic fields transported into the winds come from the circumbinary disk, either by a direct ejection as a disk wind or
as a result of the erosion with the line-driven stellar wind (see next section). We assume that the fields are 
tightly wound up since the rotational velocity is quite large at the inner disk. 
We use weak fields characterized by the ratio of the magnetic energy density to the kinetic energy density in the wind of  
\beq
\sigma = \frac{B^2}{4 \pi \rho v_{\infty}^2} = 0.005 \,\, .
\eeq

We compute three models listed in Table 1.  Model A has a wind source that is a spherical, magnetized  line-driven wind, 
with the same input equations as in Garc\'{\i}a-Segura (1997), and Models B and C are tapered, magnetized  winds (jets), with the same Gaussian functions used in Lee \& Sahai (2003) for the wind density ($\rho$) and radial
velocity ($v_{\infty}$), with a half-opening angle of $60^{\circ}$.
The wind is injected in the first two innermost
radial zones.

\section{Results}

\subsection{The Circumbinary disk}

The  expelled parts of the envelope that are gravitationally bound have sufficient angular momentum and energy 
to remain in keplerian orbits. However, due to the fact that the envelope is gaseous,
the gas particles are 
forced to collide at the equatorial regions, forming a circumbinary disk.
The thickness of the disk in the direction perpendicular
to the equatorial plane is determined by the hydrostatic equilibrium of the thermal and magnetic pressures and is directly proportional 
to the temperature of the gas and the magnetic field strength. 
The tendency of the gas to move towards the equator following an inclined orbit, although impeded by the disk,
is responsible for the triangular shape at the inner edge
(see Figure 2). 
The gas here is forced to move parallel to the
orbital plane by the pressure of the disk.
This triangular shape defines the "opening angle" of the disk. When a stellar wind collides with the inner edge of the disk, a 
shock is formed around the disk, and the shape of the shock not only depends on the Mach number, but also on the shape of the disk. 
This interaction forms the opening angle of the disk, which is the region within which the wind can freely expand.

The already formed disk is shown in Figure 2 for Models A and B, at 200 days after the simulations start from the initial
conditions ($=256.7$ days from the beginning of the CE event). 
Model A has a wind source that is a spherically symmetric, line-driven wind, while model B has a tapered flow with an imposed half-opening
angle of $60^{\circ}$. Both winds collide with the inner edge of the disk, resulting in an effective 
half-opening angle of $\sim 30^{\circ}$. The models are almost indistinguishable from one each other. 
Thus, the inner disk defines the maximum possible angle for the outflow. Larger angles are possible with time when 
radiative cooling become efficient and part of the gas has been evaporated or blown out. 

Table 2 shows that the 
mass of the envelope that remains bound is  0.517 \Mo , i.e., 
75\% of the 
envelope (Ricker \& Taam 2012). On the other hand, Table 3 shows that the  total mass remaining on the grid after 200 days for Model A  is 0.515 \Mo. This confirms that the gas is indeed bound,
and only 0.002 \Mo \, of the bound gas is able to travel farther than 1.266 AU . 
The kinetic energy in the $\phi$-direction of Model A (at 200 days) is $3.653 \times 10^{46}$ erg (Table 3),
which is $99\%$ of the total kinetic energy of Model A. This high value of the kinetic energy 
stored in the circumbinary disk is an important reservoir of energy, which could be 
``reprocessed''  and used for launching disk winds and jets. The values for the stored energy agree in order of magnitude with
those observed in HD 101584 (Olofsson et al. 2019) with ALMA (see discussion section).

The maximum magnetic field at the inner disk is $1.253  \times 10^{4}  $ gauss , which is a factor of 25 larger than the 
initial value of 500 gauss. This magnetic field is large enough to form magnetized outflows, although the precise amount that
is needed is something not answered yet by theory (Romanova \& Owocki 2015). On the other hand,
the maximum value of the magnetic field imposed at the base of the wind is only $0.328 $  gauss for Models A and B. This value
is used in order to yield $\sigma  = 0.005 \,\, $ for a mass-loss rate of  $4.19 \times 10^{-10} $ \Moy. Certainly, larger values 
for both the magnetic fields and the mass-loss rates are expected according to the inferred kinetic energy
(Bujarrabal et al. 2001; Olofsson et al. 2019), and this is a topic for a future study.

\subsection{The Formation of Highly Collimated Nebulae}

Starting from the initial conditions shown in Figure 1, we first follow the expansion of the ejected
envelope, and later on the expansion of the formed bipolar nebula by using the expanding grid technique described in detail in paper I. 
The result of the computation for Model C is shown in Figure 3 at 330 days from the initial conditions, i.e., at 386.7 days after the 
CE event. Although it is possible to follow the expansion of the nebula for many years (see paper I),
the expansion of the grid gradually leads to a lower  radial spatial resolution, and part of the circumbinary disk is eventually 
no longer resolved in the inner portion of the grid. For that reason, we show the result at 330 days, where the mass of
the disk is not totally lost. The total envelope mass of the RG was $0.69$ \Mo (Table 2), with $0.172$ \Mo \, unbound and $0.517$ \Mo \,bound, 
while the mass remaining on the grid in Model C  is $0.647$ \Mo (Table 4) at 330 days, i.e., $\sim 0.042$ \Mo \, of the bound mass is missed at 
this stage due to the expansion of the grid at the inner radial boundary. The resulting overall shape is a highly collimated, bipolar 
nebula. The shape remains quite similar after that time, since the expansion grows nearly homologously.

On the other hand, important changes occur at the inner portion of the simulation. The initial half-opening angle of $\sim 30^{\circ}$ 
changes to the imposed value of $\sim 60^{\circ}$ (see the evolution in Figure 4). This value is close to the one found by Bollen et al.(2020) 
of $\sim 76^{\circ}$.  It is far from the value obtained in Figure 2, with much higher resolution, but it is expected that
this behavior will develop in the long term, due to the decrease of mass at the inner disk associated with its erosion by the wind.  

Figure 5 shows an expanded view of Figure 3, where the self-collimation of the magnetized wind due to the magnetic pinching effect
(Begelman \& Li 1992) is clearly displayed. The high gas temperatures achieved in Model C at the convergence 
zone ($\sim 10^7$ K) are remarkable, implying X-ray emission there and likely collisional excitation of lines such as [NII] and [SII] 
in the neighborhood (see discussion section).

\subsection{The Unbound Ejected Envelope}

During a CE event, part of the orbital energy of the companion star is imparted to the envelope as kinetic energy, and
another part is converted into thermal energy, heating up the envelope. Table 5 shows that these amounts are quite similar 
in our initial conditions, where the initial kinetic energy is $3.236 \times 10^{46}$ erg, and the initial internal energy is $3.334 \times 10^{46}$ erg. Note that the kinetic energy in the $\phi$-direction is $97\%$ of the total kinetic energy, while only $0.02\%$ is in the radial component.

However, as the envelope evolves (Table 4), the kinetic energy in the r-direction has increased to $1.455 \times 10^{46}$ erg. This
increment of the energy is a result of acceleration of the gas by the thermal pressure, since the total internal energy has decreased
to a value of $3.205 \times 10^{44}$ erg by adiabatic expansion in the ejected envelope. These high values for the radial kinetic 
energy are observable in proto-planetary nebulae and cannot be achieved by the effects of radiation pressure, as radio observations have 
shown in several studies (Bujarrabal et al. 2001; Olofsson et al. 2019). In this context, the CE event behaves in a similar way to a thermal explosion. A similar behaviour was found by Iaconi et al. (2019). 

Figure 6 shows that the ejected, unbound  mass ($0.172$ \Mo) forms an excretion disk. 
 Only a very small fraction (in volume, but not in mass)  of the inner disk is bound ($\lesssim 1.3$ AU), as is shown more clearly in Figure 2 and in Table 3. Note that the bound mass is 0.517 \Mo, and the mass in Figure 2 is 0.515 \Mo. Most of the ejected mass
remains in the equatorial region, where a 
linear relation between distance and radial velocity develops, similar to an explosive event. The final 
velocities ($ 20 - 60 $ \kms) and temperatures at the head of the shock ($1,700- 7,800$ K) likely excite  molecular hydrogen, and the line at $2.12 \mu m$ 
could be observed (see discussion section). This shock is clearly
visible in the lower-right panel (temperature) of Figure 6, in yellow-orange colors.

Another important aspect is the large amount of magnetic energy stored in the excretion disk. Part of the expanding disk is 
neutral at 330 days, and some parts are partially ionized. The survival of the magnetic energy cannot be studied using our
ideal MHD scheme, and non-ideal effects such as ambipolar diffusion have to be included.

\section{Discussion}

We are still at a stage far from an understanding of the launching of the winds
in post-CE nebulae, since as we described in the introduction section, the scenario
of a ``wind soup'' from a close binary system surrounded by a circumbinary disk is 
complex and is yet to be solved. However, we have shown in this article that
the stellar magnetic field ejected with the envelope is important and can  
play a fundamental role in the launching of the winds. 
In our approximation, the field structure is toroidal throughout, so there are no poloidal fields.
The inclusion of poloidal fields in the disk could lead to the formation of a wind either via gas pressure
and magneto-centrifugal effects (sling mechanism) or magnetic pressure gradients
(spring mechanism) operating in the disk. 
For the spring mechanism the poloidal field is not necessary in principle, 
as some initial launching velocity is needed (Contopoulos 1995). This could be achieved by the action of the poloidal field (sling mechanism). A more realistic treatment of the ejection mechanism is a challenge for a
future study.

The inclusion of magnetic fields in this investigation leads to two major differences compared with the 
results reported in paper I. In particular, the resulting shape and structure of the nebula are significantly 
affected. The highly elongated bipolar shape produced in this work cannot be achieved 
without the pinching effect due to the toroidal magnetic field. The circumbinary disk acting as an obstacle to 
the wind flow is insufficient to achieve a high degree of collimation, although bipolar nebulae are obtained in both cases. Another major difference is the production of jets in the polar regions in the magnetized case. 
Here, this more extreme feature arises from the pinching effect caused by the increment
of the magnetic pressure close to the polar axis. Begelman \& Li (1992) already 
showed that the magnetic pressure within a shocked bubble acquires a cylindrical structure,
being much larger (dynamically dominant) close to the axis, without variation in the $z$ (vertical)
direction. The formed jets, as the one observed in the polar regions of Figure 3, 
are dynamically structured also as a linear relation between distance and radial velocity
(Figure 7) in the velocity profile. This behavior was
observed, for example, by  Bryce et al. (1997) in MyCn 18 (the Hour Glass nebula).
This is similar to the velocity law produced in the plasma gun scenario (see Figure 4 in the work of Contopoulos 1995). Both of these results were already discussed in Garc\'{\i}a-Segura et al. (1999). 
The difference here is the origin of the magnetic field. While the magnetic
field in this study is from the circumbinary disk (previously inner part of
the stellar envelope), which is highly rotating  at $235 \kms$ (Table 3), 
the magnetic field in Garc\'{\i}a-Segura et al. (1999) is from the surface of the PN central star. 
The PN central star rotates very slowly in the case of single stars (Garc\'{\i}a-Segura et al. 2014), 
but not in the case of a CE (Ricker \& Taam 2012). With this in mind, neither of the models
can be discarded yet. However, line-driven winds from PN central stars cannot provide the 
necessary amount of kinetic energy observed in proto-PN nebulae (Bujarrabal et al 2001). This
fact makes circumbinary  disks
a better candidate to explain proto-PN.
As we commented earlier, the exact solution for the magnetic launching of the winds or jets is not well  understood, and could be produced by the disk itself, by accretion onto
the binary system, or a combination of both. This will be the topic for our
next article.

The lifetime of the circumbinary disk is limited, and so are the effects from the magnetic
field. On the basis of this reasoning, realistic modelling of a PN descended
from a CE has to take into account the magnetic nature of the winds in the first part 
of the computations (proto-PN stage),  which progressively evolve toward a non-magnetic, 
line-driven wind as in paper I (PN stage). This will be the scope of a future study.

There are a large number of observations that can be discussed on the light of these new calculations. We 
focus here on some of them.

Olofsson et al. (2019)  reported measurements of the kinetic energy for the molecular gas of
$7 \times  10^{45}$ [D/1 kpc]$^2$ erg, or $2 \times 10^{45}$ erg  (low-$L_*$) and $ 2 \times 10^{46} $ erg (high-$L_*$) 
in the object HD 101584. This object, according to the above study, could be  a nebula
associated with a post-RG star (low-$L_*$ case) or with a post-AGB star (high-$L_*$ case),
although the first case is preferred by the authors. 
Our reported values for the kinetic energy are able to attain such high values. For example, 
the kinetic energy in the $\phi$-direction (Model A) is $3.653 \times 10^{46} $ erg (Table 3), 
while the kinetic energy in the  $r$-direction  (Model C)   is  $1.455  \times 10^{46} $ erg (Table 4).
Although our model was not tailored to HD 101584, the agreement is remarkable in many
aspects, such as the shape (their Figure 18)  and  the self-collimation of the winds. Their sketch (Figure 4) is quite similar to our Figure 4.
The temperatures in our Figures 3 and 5 are probably too high to explain
the molecular observations. However, they provide an indication
of where the SiO should be observed in the high velocity outflow (HVO) 
and the CO in the hour glass structure (HGS), for a model with lower
outflow velocities, in very good morphological agreement. In addition, the equatorial density 
enhancement (EDE) can be straightforwardly associated with the unbound ejected envelope. 

The temperatures achieved in our Figure 3 and 5 of $10^{6}-10^{7}$ K are more 
successful in explaining the PN  Hen 3-1475. The gas at these high temperatures is able to emit X-ray radiation, 
and the hot spot at the convergence zone of the wind with higher densities is the 
ideal location for their detection. We note that this hot spot is the only place where X-rays were reported in Hen 3-1475 (Sahai et al 2003). The self-absorption of X-rays by gas within the nebula considerably reduces the detection of soft X-rays, and this provides a plausible explanation for the lack of detection of X-ray emission in the other parts of the nebula, given the sensitivity of Chandra.
Finally, the emission of  [NII] and [SII] in Hen 3-1475 (see Fang et al. 2018 and references therein)
is related to shocked gas of moderate temperatures, and it is likely that they arise from the jets and the forward shock of the bipolar nebula. However, the jets in Hen 3-1475 are broken into bullets of gas, 
which are a natural consequence of neck instabilities (Jackson 1962).

Another observation related to the results of this study is the detection of
$H_2$  ($2.12 \mu$m ) in the proto-PN  CRL 2688 (the Egg Nebula) (Sahai et al. 1998).
Specifically, the emission that appears to be related to the equatorial region. 
Figure 6 shows that the unbound ejected envelope in the equatorial region is not 
photoionized, and the temperature at the head of the shock reaches $\sim 7,800$ K.
This temperature is sufficiently high to lead to the excitation of molecular hydrogen
(Burton et al. 1992). Thus, 
according to our computations, the observation of shocked molecular hydrogen
in the equatorial regions of CRL 2688 is a direct confirmation that the 
nebula was formed in a CE scenario, which is the only one that can eject such 
an amount of gas in the equator at moderate velocity. Note that our computations
are axisymmetric, but the original 3D FLASH computation (Ricker \& Taam 2012) does
not eject the equatorial gas in all the $\phi$-directions in the same manner. This
likely provides an explanation for the asymmetric distribution of shocked molecular hydrogen in CRL 2688 around the central star. 

\section{Conclusions}

The predicted strengths of the magnetic fields inside red giant stars based on models calculated using stellar evolution codes can produce fields sufficient to produce disk winds at the self-regulated phases after CE events, or at least, be an important seed for the generation of even stronger fields by a dynamo. We find that the 
fields are amplified by a factor of 25 due to gas compression on the recently formed circumstellar disk.

The kinetic energy stored in the circumbinary disk ($\sim 10^{46}$ erg) is probably sufficient
to account for the observed values of kinetic energies in proto-PN outflows, once this energy
is released and converted into radial kinetic energy by a magneto-centrifugal scenario. 

We prove that, independently of the type of wind source, the resulting winds behave as ``tapered'' flows 
during the initial phases of proto-PNe, in agreement with previous studies in the literature
(Lee \& Sahai 2003; Balick et al. 2019).  Although the calculations presented
by Zou et al.\ (2019) also show a similar result, we showed in paper I that non-magnetic
spherical winds cannot be collimated to form tapered flows or jets. This apparent 
discrepancy does not result from
the hydrodynamical simulation by Zou et al.\ (2019) itself, using AstroBEAR,
but from the 
initial conditions extracted from the SPH simulation using the code PHANTOM,
which use adiabatic conditions up to 14 years after the ejection. 
Note that this simulation was aimed at calculating
the final separation of the binary and the efficiency of the
ejection, but not resolving the shape and dynamics of the ejecta, where radiative cooling 
is intense after a short, initial adiabatic episode. The small
opening angle and the motion of the ejected gas towards the polar regions, as seen in their Figure 1,  is just a consequence of the adiabatic conditions. 

Small values of $\sigma$ ($= 0.005 $) are sufficient to explain wind self-collimation in proto-PN. 
Finally, it is found that X-ray emission and blobs of shocked gas are a consequence of the pinching effect 
from the toroidal magnetic fields.

\acknowledgments

We thank Michael L. Norman and the Laboratory for Computational
Astrophysics for the use of ZEUS-3D. The computations
were performed at the Instituto de Astronom\'{\i}a-UNAM at  Ensenada.
G.G.-S. is partially supported by CONACyT grant 178253.
Partial support
for this work has been provided by NSF through grants
AST-0200876 and AST-0703950. FLASH Computations were carried
out using NSF Teragrid resources at the National Center for 
Supercomputing Applications (NCSA) and the Texas Advanced
Computing Center (TACC) under allocations TG-AST040024
and TG-AST040034N. P.M.R. acknowledges support from NSF under AST 14-13367 and
the Kavli Institute for Theoretical Physics, where some of this work was performed 
with funding by NSF under grant PHY05-51164. FLASH
was developed and is maintained largely by the DOE-supported
Flash Center for Computational Science at the University of
Chicago

{\bf \software{ZEUS-3D (version 3.4; Stone \& Norman 1992, Clarke 1996)}}
{\bf \software{Binary Evolution Code (Petrovic et al. 2005; Yoon et al. 2006)}}
{\bf \software{FLASH (2.4; Fryxell et al. 2000)}}



\clearpage

\begin{table}
\begin{center}
\caption{Numerical models}
\begin{tabular}{lccccccr}
\tableline\tableline
Model &  Resolution & Type  & Grid & Figures \\
\tableline
A & $200 \times 200$ & Spherical & Fixed     & 2           \\
B & $200 \times 200$ & Tapered   & Fixed     & 2           \\
C & $400 \times 400$ & Tapered   & Expanding & 3,4,5,6,7     \\
\tableline
\end{tabular}
\end{center}
\end{table}

\begin{table}
\begin{center}
\caption{Values in Ricker \& Taam (2012)}
\begin{tabular}{lr}
\tableline\tableline
Variable &  Value (\Mo)   \\
\tableline
Total mass of primary       &          1.05 \\
Total mass of secondary     &          0.6  \\
Core mass of primary        &         0.36  \\
Envelope mass of primary    &          0.69 \\
Total ejected envelope mass          & (25\%) 0.172  \\
Total bound envelope  mass            & (75\%) 0.517  \\
\tableline
\end{tabular}
\end{center}
\end{table}

\begin{table}
\begin{center}
\caption{Model A at time 200 days}
\begin{tabular}{lr}
\tableline\tableline
Variable &  Value (cgs)  \\
\tableline
Total mass in the grid & (0.515  \Mo) $1.023 \times 10^{33} $ \\
Central point mass   & (0.96 \Mo)  $1.909 \times 10^{33}$ \\
Kinetic energy in $\phi$-direction         & $3.653 \times 10^{46} $   \\
Total kinetic energy                       & $3.657 \times 10^{46} $   \\
Gravitational potential energy             & $-6.845 \times 10^{45} $   \\
Magnetic energy                            & $1.095  \times 10^{44 }$   \\
Maximum density                            & $1.144  \times 10^{-3} $  \\
Maximum velocity in $\phi$-direction       & $235    \times 10^{5}  $  \\
Maximum magnetic field                     & $1.253  \times 10^{4}  $  \\
Maximum magnetic field at base of wind     & $0.328      $  \\
\tableline
\end{tabular}
\end{center}
\end{table}

\begin{table}
\begin{center}
\caption{Model C  at time 330 days}
\begin{tabular}{lr}
\tableline\tableline
Variable &  Value (cgs)  \\
\tableline
Total mass in the grid &    (0.647  \Mo) $1.287  \times 10^{33} $  \\
Central point mass     &    (0.96 \Mo)   $1.909 \times 10^{33}  $  \\
Kinetic energy in  r-direction      &    $1.455  \times 10^{46} $  \\
Kinetic energy in  $\theta$-direction &  $5.100  \times 10^{44} $  \\
Kinetic energy in  $\phi$-direction &    $1.749  \times 10^{46} $  \\
Total kinetic energy                &    $3.255  \times 10^{46} $  \\  
Total internal energy               &    $3.205  \times 10^{44} $  \\
Gravitational potential energy      &    $-6.601 \times 10^{44} $  \\
Magnetic energy                     &    $2.422  \times 10^{43} $  \\
Maximum density                     &    $2.935  \times 10^{-7}$   \\ 
Maximum magnetic field             &     $7.402   \times 10^{1} $  \\
Maximum magnetic field at base of wind & $3.017  \times 10^{-2} $  \\
\tableline
\end{tabular}
\end{center}
\end{table}

\begin{table}
\begin{center}
\caption{Initial conditions for Model C}
\begin{tabular}{lr}
\tableline\tableline
Variable &  Value (cgs)  \\
\tableline
 Total mass in the grid &    ( 0.689  \Mo) $ 1.3705  \times 10^{33}$ \\ 
Kinetic energy in  $r$-direction      & $9.090  \times 10^{44} $  \\
Kinetic energy in  $\phi$-direction & $3.145  \times 10^{46} $  \\
Total kinetic energy                & $3.236  \times 10^{46} $  \\  
Total internal energy               & $3.334  \times 10^{46} $  \\
Gravitational potential energy      & $-8.977 \times 10^{45} $  \\
Magnetic energy                     & $6.265  \times 10^{43} $  \\
\tableline
\end{tabular}
\end{center}
\end{table}

\clearpage

\begin{figure}
\epsscale{1.20}
\plotone{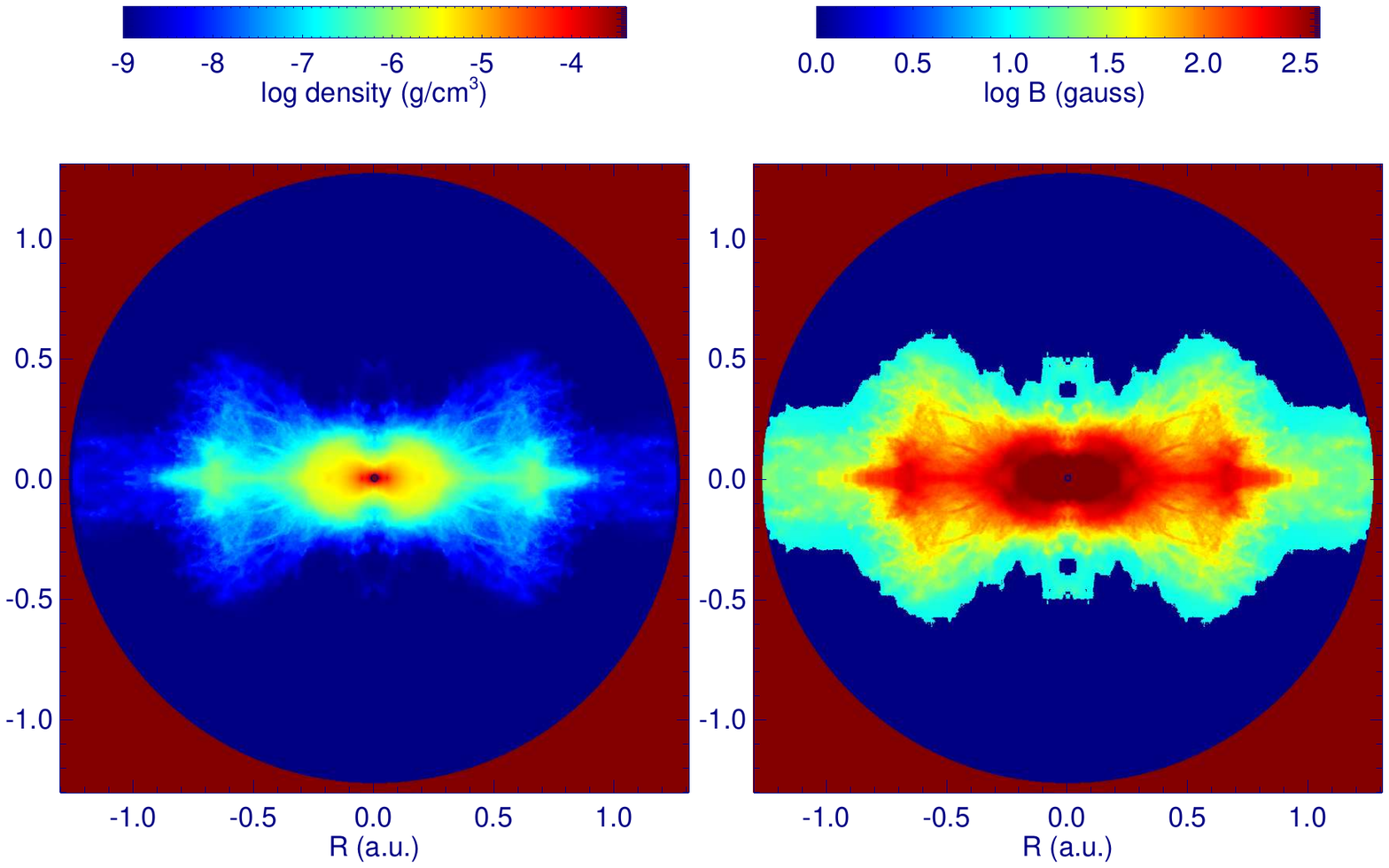}
\caption{Gas density and magnetic field of the ejected envelope (initial conditions).
The equatorial plane is perpendicular to the figure and  horizontally oriented, passing 
through the center. The symmetry 
axis is an imaginary, vertical line that passes through the center. Total mass is 0.689 \Mo . The magnetic field is toroidal, perpendicularly oriented to the figure. The binary at the center has a period of $\sim 5$ days and a separation of $\sim 0.04 $ AU .}
\label{f1}
\end{figure}

\clearpage

\begin{figure}
\epsscale{1.20}
\plotone{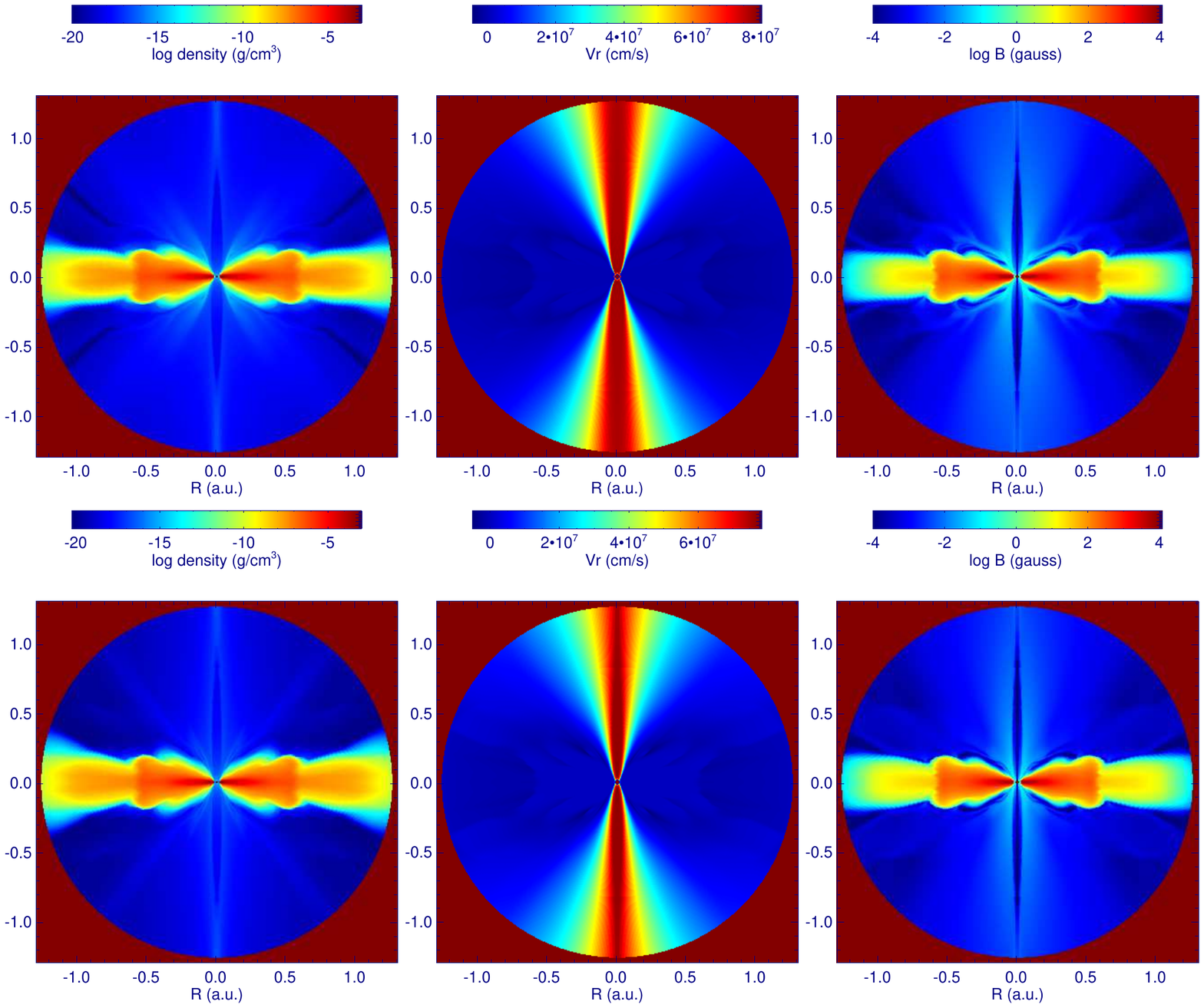}
\caption{Gas density, radial velocity and magnetic field of Model A (top) and Model B (bottom) at  time 200 days.}
\label{f2}
\end{figure}

\clearpage

\clearpage

\begin{figure}
\epsscale{1.20}
\plotone{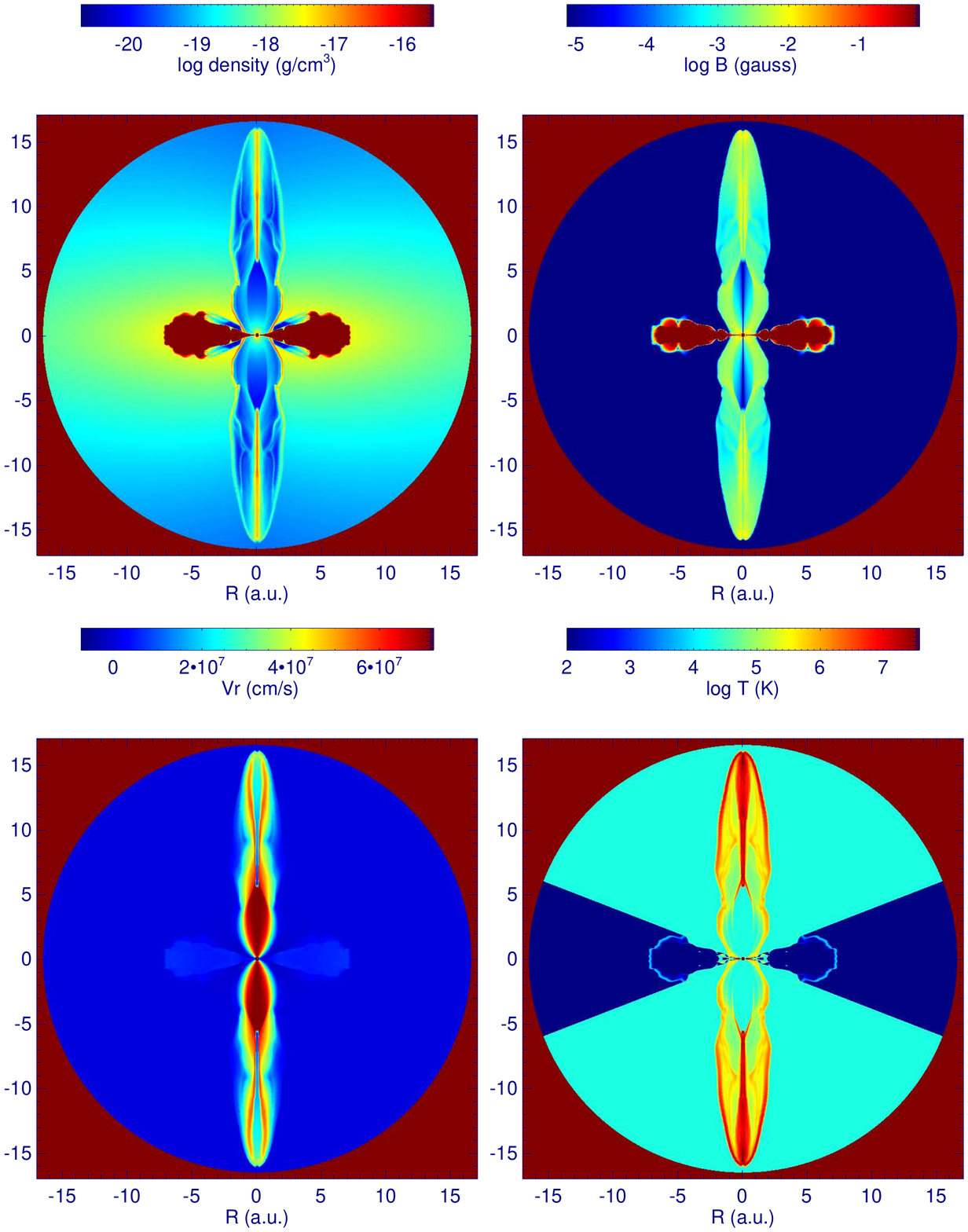}
\caption{Gas density, magnetic field, radial velocity and temperature of Model C  at time 330 days.}
\label{f3}
\end{figure}

\clearpage

\begin{figure}
\epsscale{1.20}
\plotone{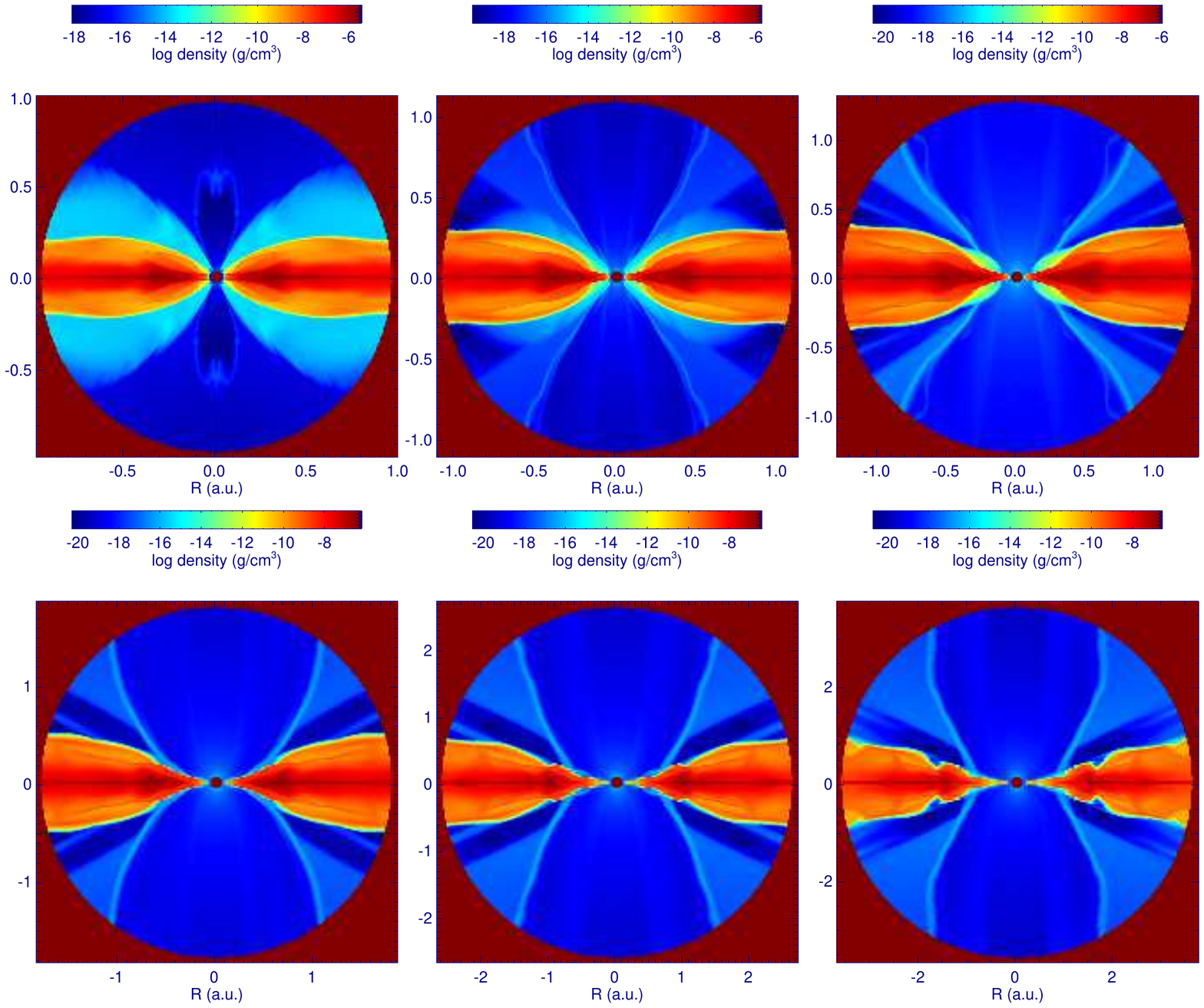}
\caption{Model C. Evolution of the inner disk (left to right) 
at times  120, 140, 160, 180, 200, and 220 days. Only the first 100 radial grid zones
are displayed.}
\label{f4}
\end{figure}

\clearpage

\begin{figure}
\epsscale{1.20}
\plotone{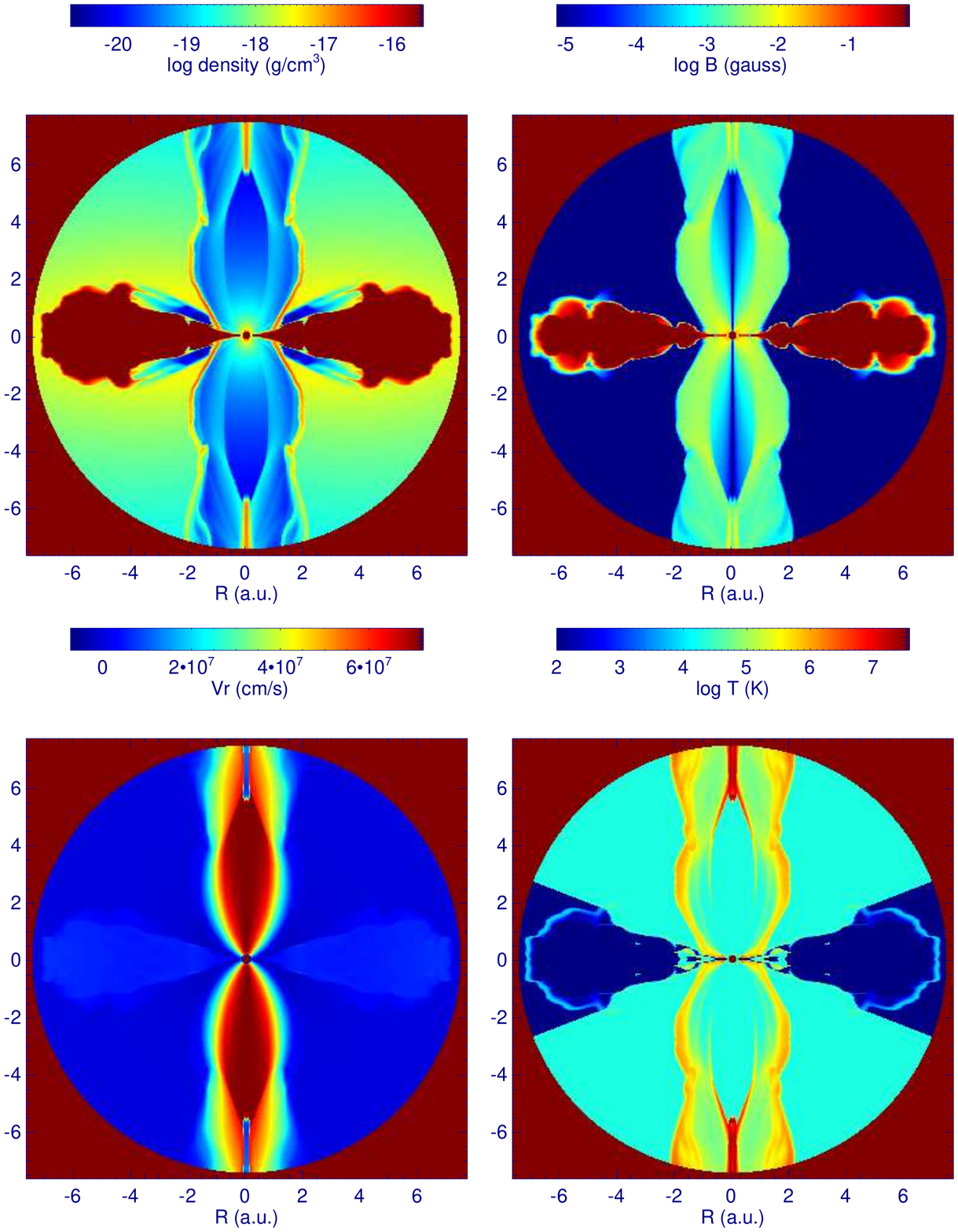}
\caption{Zoom of gas density, magnetic field, radial velocity and temperature of Model C  at time 330 days, with
the color scale optimized for the collimated outflow.}
\label{f5}
\end{figure}

\clearpage

\begin{figure}
\epsscale{1.20}
\plotone{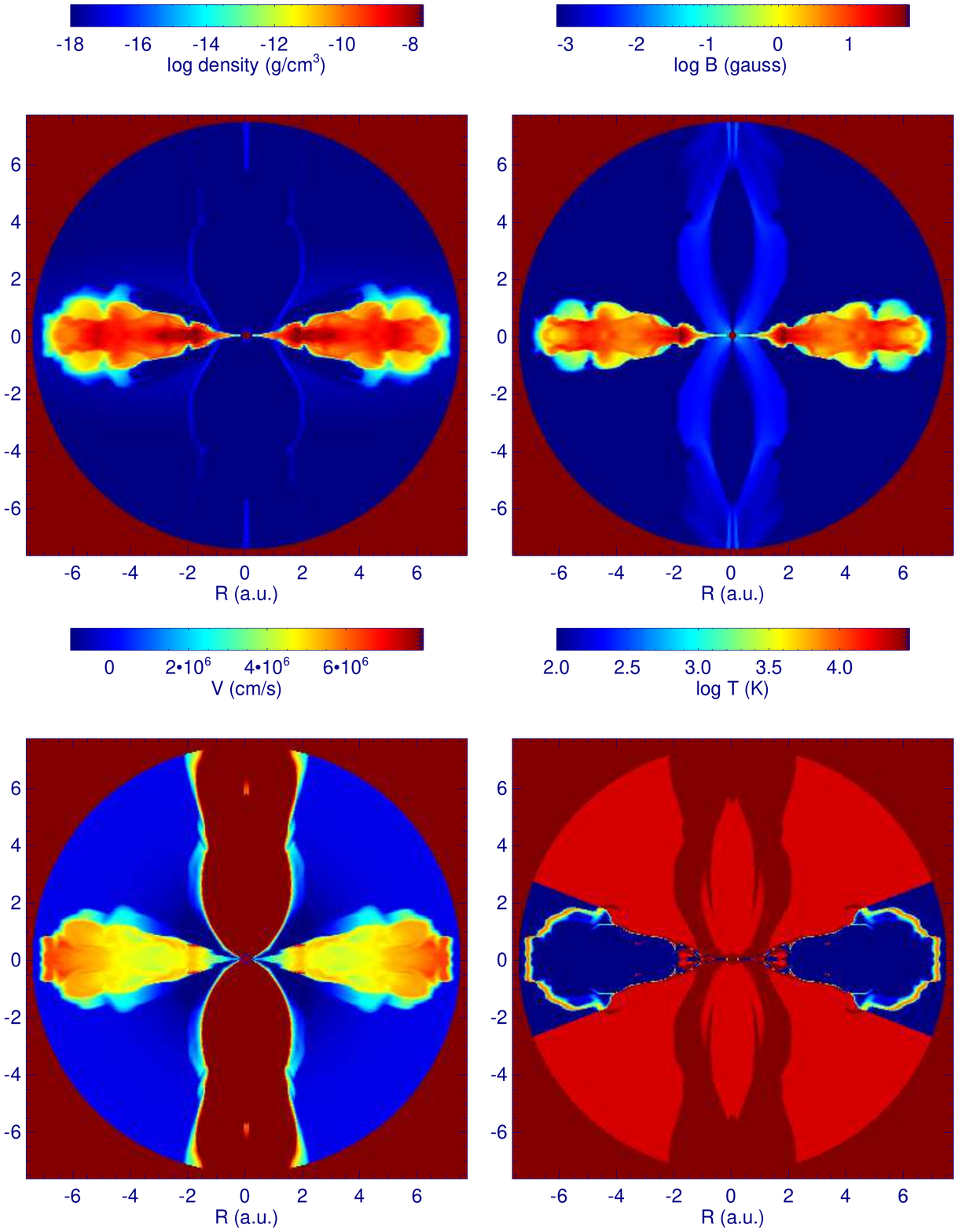}
\caption{Zoom of gas density, magnetic field, radial velocity and temperature of Model C  at time 330 days, with
the color scale optimized for the ejected envelope.}
\label{f6}
\end{figure}

\clearpage

\begin{figure}
\epsscale{1.20}
\plotone{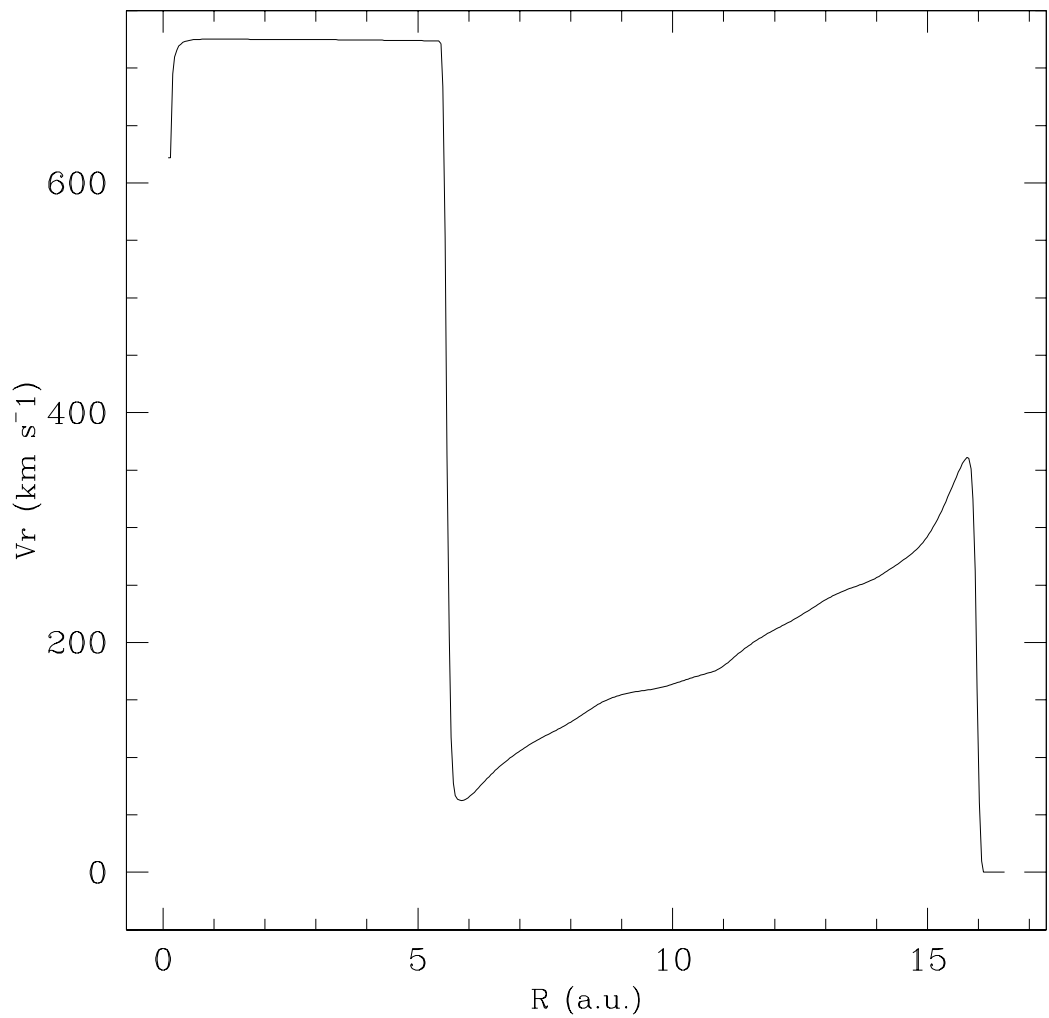}
\caption{Radial velocity at $\theta=0$  (polar axis) from Model C at time 330 days. }
\label{f7}
\end{figure}

\clearpage
\end{document}